\documentclass[twocolumn,prl,showpacs,superscriptaddress]{revtex4} 

\usepackage{graphicx}
\usepackage{dcolumn}
\usepackage{bm}
 
\begin{document} 
\def\k {{\bf k}} 
\def\r {{\bf r}} 
\def\R {{\bf R}} 
\def\H {{\bf H}} 
 
\preprint{Taraskin et al.} 
 
\title{Spatial decay of the single-particle density matrix in insulators: \\ 
 analytic results in two and 
three dimensions}

\author{S.~N.~Taraskin} 
 \email{snt1000@cus.cam.ac.uk} 
\affiliation{Department of Chemistry, University of Cambridge, 
             Lensfield Road, Cambridge CB2 1EW, UK} 
 
\author{D.~A.~Drabold}%
 \email{drabold@ohio.edu} 
 \altaffiliation[Work implemented while on leave at~]{Trinity College, Cambridge} 
\affiliation{Department of Physics and Astronomy, Ohio 
University, Athens, Ohio 45701 } 
 
\author{S.~R.~Elliott} 
\email{sre1@cus.cam.ac.uk} \affiliation{Department of Chemistry, 
University of Cambridge, 
             Lensfield Road, Cambridge CB2 1EW, UK} 
 
\date{\today}
 
\begin{abstract} 
Analytic results for the asymptotic decay of the electron density 
matrix in insulators have been obtained in all three dimensions 
($D=1 - 3$) for a tight-binding model defined on a simple cubic 
lattice. The anisotropic decay length is shown to be dependent on 
the energy parameters of the model. The existence of the 
power-law prefactor, $\propto r^{-D/2}$, is demonstrated. 
\end{abstract} 
 
 
 
\maketitle 
 
 
In a phenomenological classical approach to atomic  dynamics,  a 
very local picture of interatomic interactions is often highly 
appropriate. The simplest example is the use of Keating 
''springs'' to describe small atomic oscillations in solids 
\cite{Keating_66}. In quantum mechanics (QM), the situation is 
superficially quite different. If $\psi_{\text I}$ are 
single-particle wave functions (for example, the Kohn-Sham 
orbitals of density-functional theory), then the  electronic 
contributions to the total energy, forces or the dynamical matrix 
can all be obtained from the single-particle density matrix (DM): 
$\rho({\bf r},{\bf r'})= \sum_{\text{occ}} \psi_{\text I}^*({\bf 
r}) \psi_{\text I}({\bf r'})$ \cite{Goedecker_99:review}.  For a 
condensed-matter system, virtually all of the Hamiltonian 
eigenstates $\psi_{\text I}$ are extended and oscillatory 
throughout the volume of the system, except some of the states 
near the band edges in the case of disordered systems. To the 
extent that $\rho({\bf r},{\bf r'})$ is built from objects that 
are extended in real space, it is a nontrivial feature of the 
quantum mechanics of extended systems that in fact the DM can be 
localized, even exponentially so in insulators 
\cite{Kohn_73,Rehr_74}. 
 
The possibility of a local formulation of QM goes back at least to 
Wannier \cite{Wannier_37}, whose ''Wannier functions'' decay in 
real space. 
Kohn \cite{Kohn_96} extended this work, and emphasized 
the principle of ''nearsightedness'', that is the dependence of 
local properties, such as forces,  on the local environment. 
The possibility of real-space local formulations of QM has inspired 
vigorous activity on efficient ''order-N''  methods for 
electronic structure \cite{Goedecker_99:review}. 

In early pioneering work, Kohn \cite{Kohn_59} showed that the DM 
and Wannier functions for a 1D centrosymmetric two-band model 
decay exponentially in insulators: $\rho(\r,0) \propto \exp(- 
\lambda |\r|$), where $\lambda \propto E_{\text{gap}}^{1/2}$ 
\cite{Kohn_59}, although Ismail-Beigi and Arias 
\cite{IsmailBeigi_99} find instead that $\lambda \propto 
E_{\text{gap}}$ as $E_{\text{gap}} \to 0$. Des Cloizeaux proved 
that the DM decays exponentially quite generally in insulators 
\cite{Cloizeaux_64}. Recently, He and Vanderbilt \cite{He_01} 
demonstrated  that in $1D$ there is in fact a power-law 
prefactor  $\propto r^{-1/2}$, and that the characteristic 
inverse decay length, $\lambda$,  of both the DM and Wannier 
functions is universal, although the prefactor is different in 
each case. 
 
In this Letter, we report analytic results for the spatial decay 
of the DM for a two-band tight-binding (TB)  Hamiltonian for all 
three dimensions, $D=1 - 3$. The Hamiltonian we use is similar in 
spirit to that of  Harrison's bond orbital model 
\cite{Harrison_80} and is the minimal model which contains the 
basic features of an insulator. Both the power-law prefactor and 
the exponential spatial decay of the DM are found in the 
asymptotic regime. Thus, we reproduce the results of 
Refs.~\cite{Kohn_59,He_01}, and extend them to  2D and 3D. We 
derive the dependence of the inverse decay length, $\lambda$, in 
terms of the fundamental energy parameters of the model. We 
further show how, in certain cases,  the exponential localization 
arises from the truncation of the sum defining $\rho$. Precise 
numerical results support the analytic asymptotic behaviour  in 
all three dimensions. 
 
Our model is of centrosymmetric nearest-neighbor TB form on a 
simple cubic (s.c.) lattice described by the Hamiltonian: 
\begin{equation} 
{\hat {\bf H}}= \sum_{i\mu}\varepsilon_\mu |i\mu\rangle\langle 
i\mu| + \sum_{i\mu,j(i)\mu'}t_{\mu\mu'}|i\mu\rangle\langle j\mu'| 
\ . \label{e1} 
\end{equation} 
Here, each site $i$ is characterized by two  bare orthogonal 
electronic states, $|i\mu\rangle$ ($\mu =1,2$) with bare energies 
$\varepsilon_\mu$ ($\Delta\varepsilon =\varepsilon_1 - 
\varepsilon_2 > 0$ for definiteness), which together form the 
orthonormal bare basis set. These may be interpreted as bonding 
and antibonding states centered at each lattice site. 
The hopping  integrals, $t_{\mu\mu}$, between similar states (the 
same $\mu$) on different nearest-neighbour sites $j$ (to $i$) 
transform the on-site energy levels into  bands, while the 
hopping integrals, $t_{\mu\mu'}$, between different states ($\mu 
\ne \mu'$) on different sites are responsible for inter-band 
hybridization. 
 For this model, the eigenstates $|\k,\gamma\rangle$ ($\k$ is the 
 wavevector and $\gamma=1,2$ indexes the bands) 
  and dispersion are analytically available 
and the spectrum exhibits the 
 main features of a semiconductor/insulator 
(i.e. two bands separated by a 
 gap for a certain range of parameters).

%
%
%
%
%
%
%
%
%
\begin{figure} 
\centerline{\includegraphics[width=6cm,angle=270]{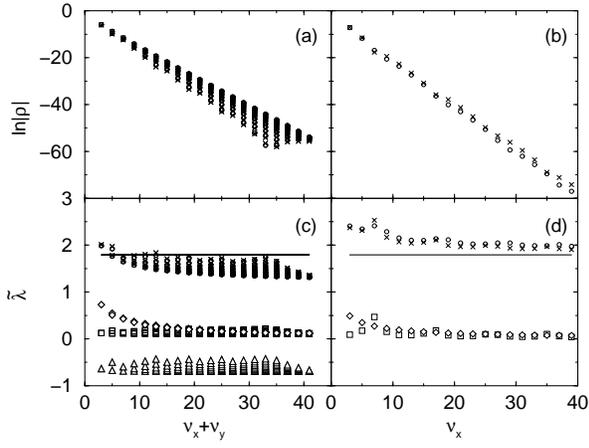}} \caption{ 
The dependence on distance of the logarithm of the DM 
$|\ln\rho_{\nu_\alpha}|$ (a,b) and of the effective 
inverse decay length, ${\tilde\lambda} = -\ln|\rho_{\nu_\alpha}|/ \sigma $ (c,d)  in the $2D$ case 
(the 
prefactor is included in the definition of ${\tilde\lambda}$ for 
easier comparison of analytical results with numerical ones). 
In (a) and 
(c), the DM has been calculated on the lattice sites situated 
around the main diagonal $[11]$ and characterized by polar 
angles $\phi \in [10^0, 80^0]$, while in (b) and (d), $\nu_{\rm 
y}=0$ and $\nu_{\rm x}$ varies (in the $[10]$ direction). The 
circles represent  exact numerical results obtained  using 
Eq.~(\protect\ref{e2}) or Eq.~(\protect\ref{e3}), while the 
crosses are the results of the analytical 
expressions~(\protect\ref{e4})-(\protect\ref{e6}) (they are 
practically indistinguishable in (a) and (c)). In (c),  the 
different analytical contributions to the effective inverse decay 
length are shown by the solid line (the $\ln(2A)$-term in 
Eq.~(\protect\ref{e5})), squares (the second term in 
Eq.~(\protect\ref{e5})), triangles (the sum of the last two  terms 
in Eq.~(\protect\ref{e5})) and diamonds (the prefactor in 
Eq.~(\protect\ref{e4})). In (d), 
 the solid line is the $\ln(2A)$-contribution to 
the effective inverse decay length (see Eq.~(\protect\ref{e6})), 
the squares represent  the contribution of $(-\ln|J_{\nu_{\rm 
y}}(\nu_{\rm x}/A)|/\nu_{\rm x})$ to ${\tilde\lambda}$ and the 
diamonds signify the prefactor in Eq.~(\protect\ref{e6}). } 
\label{f1} 
\end{figure}

 We have written the density operator for the Hamiltonian given by 
 Eq.~(\ref{e1}), first, in the eigenvector basis, $|\k,\gamma\rangle$, then 
 expanded the eigenstates in the site basis $|i\mu\rangle$ and evaluated 
 the density matrix elements, 
$\rho_{i\mu,j\mu'}$ in the site basis. For fixed site indices $i$ 
and $j$, the DM contains four elements, which, in the case of a 
semiconductor at zero temperature, obey the following relations: $ 
\sum_\mu\rho_{i\mu,j\mu} = \delta_{ij} $ and $ \rho_{i1,j2}= 
\rho_{i2,j1} $. Below we investigate the decay properties only 
for the off-diagonal elements, $ \rho_{i1,j2}$ (the analysis and 
results 
 for the diagonal elements 
$\rho_{i\mu,j\mu}$ are very similar to those for $\rho_{i1,j2}$ 
presented 
 below). 
  The expression for $\rho_{i1,j2}\equiv \rho(\r_{ij})$ 
 can be conveniently represented via an integral in reciprocal space over 
 the first Brillouin zone: 
\begin{equation} 
\rho(\r_{ij}) = \frac{-1}{2(2\pi)^D} 
\int\dots\int\limits_{-\pi}^{\pi} d\k \frac{ e^{\text{i} \k \cdot 
\r_{ij}} S_\k} 
         {(A_{\k}^2+S_\k^2)^{1/2}}  \ , 
\label{e2} 
\end{equation} 
where $\r_{ij}\{a\nu_x, a\nu_y, a\nu_z\} = \r_j - \r_i$ is the 
connection vector (with the unit-cell parameter $a$ taken 
hereafter to be unity,  and $\nu_\alpha $ (e.g. $\alpha = x,y,z$ 
in $3D$) being an integer), $ S_\k=(1/2)\sum_{j(i)}\exp{({\rm 
i}\k \cdot \r_{ij})} $ is  the structure factor and $ A_{\k}= 
(\Delta\varepsilon +2(t_{11}-t_{22}) S_\k )/4t_{12} $ is the only 
energy  parameter of the model which, for the symmetric case 
($t_{11}=t_{22} \equiv t $), analyzed below for simplicity, is 
$\k$-independent i.e. $A_\k \equiv A=\Delta\varepsilon/4t_{12}$. 
In the model without interband hybridization ($t_{12}=0$ or $A\to 
\infty$), the DM given by Eq.~(\ref{e2}) is obviously 
$\rho_{ij}=0$. In order to study the spatial decay of the density 
matrix, we should take the integral in Eq.~(\ref{e2}) and 
investigate its dependence on the lattice indices $\nu_\alpha$. 
This can be done analytically in the asymptotic regime for all or 
some $\nu_\alpha \gg 1$, if the energy parameter of the model 
obeys the inequality, $A>D$, which is the case for weak interband 
hybridization in  semiconductors. 
 
 Eq.~(\ref{e2}) has been derived without 
making use of the particular symmetry type of the underlying 
lattice, and is valid for any primitive lattice (one atom per 
unit cell). The structure factor, $S_\k$, reflects the differences 
between various lattices. The DM given by Eq.~(\ref{e2}) can be 
easily calculated numerically for any reasonable choice of 
parameters, and the results of such calculations support the main 
conclusions derived below analytically for the s.c.  
lattice,  subject to some additional restrictions (symmetric 
case, $t_{11}=t_{22}$, and weak hybridization, $A>D$). 
 
\begin{figure} 
\centerline{\includegraphics[width=6cm,angle=270]{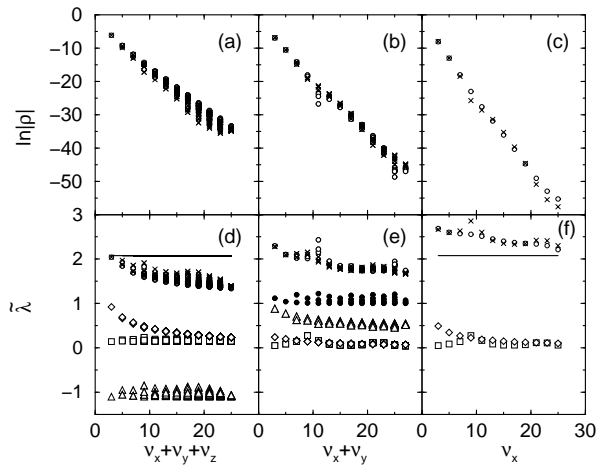}} \caption{ 
The dependence on distance of 
$|\ln\rho_{\nu_\alpha}|$ (a-c) and of  ${\tilde\lambda} = 
-\ln|\rho_{\nu_\alpha}|/ \sigma $ (d-f) in 
the $3D$ case. In (a) and (d), the DM has been calculated on the 
lattice sites situated around the main diagonal $[111]$ for 
solid angles defined by the polar angle $\phi \in [20^0, 70^0]$ 
and azimuthal angle $\theta\in[20^0, 70^0]$; in (b) and (e), 
$\nu_{\rm z}=0$ and $\nu_{\rm x}$ and $\nu_{\rm y}$ vary around 
the plane diagonal and are characterized by  the polar  angle 
$\phi \in [20^0, 70^0]$ (around the $[110]$ direction); and in 
(c) and (f), $\nu_{\rm z}=\nu_{\rm y}=0$ and $\nu_{\rm x}$ 
varies  (in the $[100]$ direction). All the symbols have the 
same meaning as in Fig.~\protect\ref{f1} except in (e), where the 
solid circles represent the contribution to $\tilde\lambda$ due to 
the exponential term  in Eq.~(\protect\ref{e7}), the squares and 
triangles 
 are due to $J_{\nu_{\rm z}}$ and 
$J_{\nu_{\rm x}+\nu_{\rm y}}$, respectively, while the diamonds 
signify the prefactor in Eq.~(\protect\ref{e7}). } \label{f2} 
\end{figure}

To proceed with the evaluation of the DM given by Eq.~(\ref{e2}), 
first we expand in a series the denominator of the integrand 
($S_\k/A < 1$, if $A>D$), separate the variables and then 
evaluate all $D$ integrals making use of the orthogonality of the 
$\cos(k_\alpha\nu_\alpha)$-functions involved. This results in 
the following general exact form for the DM for all three 
dimensions: 
\begin{equation} 
\rho_{\nu_\alpha}= \frac{(-1)^{\overline\nu}      } 
     {(4A)^{2{\overline\nu}+1} } 
\sum_{k=0}^{\infty} (-1)^k \left[ \frac{ (2k')!      } 
     { (4A)^k (k')! } 
\right]^2 
 (2k'+1) \Sigma_D~, 
\label{e3} 
\end{equation} 
if $\sigma \equiv \sum_\alpha^D \nu_\alpha$ is odd, and zero 
otherwise. 
 Here $k'= 
k+{\overline\nu}\equiv k + (\sigma-1)/2$, 
$\Sigma_1=1/[k!(k+\nu_{\rm x})!]$ and $\Sigma_{2,3}$ are 
hypergeometric finite sums, the one for $2D$, $ \Sigma_2= 
\sum_{m=0}^k 1/[m!(k-m)!(m+\nu_{\rm x})! (k-m+\nu_{\rm y})!] $ 
being expressible  in closed form \cite{Watson_66}, but that for 
$3D$, $ \Sigma_3 = \sum_{m=0}^k (2m+\nu_{\rm x} +\nu_{\rm y})! 
/[m!(k-m)! (m+\nu_{\rm x})! (m+\nu_{\rm y})! (m+\nu_{\rm x} 
+\nu_{\rm y})! (k-m+\nu_{\rm z})!] $, cannot so be written, in 
principle \cite{Petkovsek_97:book}. The orthogonality of the 
$\cos(k_\alpha\nu_\alpha)$-functions, being a consequence of the 
Fermi level lying in the gap (only in this case are the upper 
limits of the integrals in Eq.~(\ref{e2}) equal to $\pi$), 
 is an important feature in deriving Eq.~(\ref{e3}). 
The original series is truncated for small indices which results 
in the appearance of the exponential factor, $(4A)^{ 
{-2{\overline\nu}-1} }$, before the sum in Eq.~(\ref{e3}). This 
factor then appreciably contributes to the energy ($A$)-dependent 
part of the decay length of the DM 
(although see Fig.~\ref{f2}(e) 
where this is not the case). 
 
%
%
%
%
%
%
%
%
%
%
%
 
%
%
%
%
%
%
%
%
%
%
%
%
 
\begin{figure} 
\centerline{\includegraphics[width=6cm,angle=270]{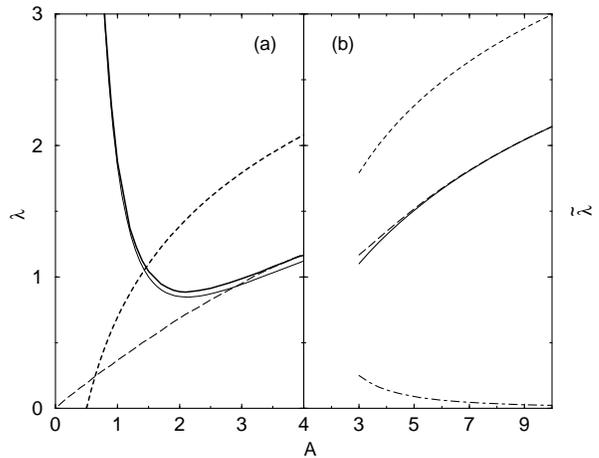}} 
\caption{ 
The dependence of the inverse decay length $\lambda$ (a) 
and the effective inverse decay length $\tilde\lambda$ (b) on the 
energy-dependent parameter $A$ of the 
model in $3D$. 
The data in (a) given by the thick 
solid and long dashed lines were obtained by linear regression from 
the slope of $\ln|\rho|$ evaluated numerically using 
Eqs.~(\protect\ref{e4}) and  (\protect\ref{e2}) for the 
$[111]$ direction, respectively. The thin solid line in (a) 
represents Eq.~(\protect\ref{e5}) (the 
difference between the thin and thick solid lines is due to the 
prefactor in Eq.~(\protect\ref{e4})) while the short dashed line shows 
the $\ln(2A)$-contribution to $\lambda$. 
The solid 
line in (b) represents the analytical result according to 
Eqs.~(\protect\ref{e4})-(\protect\ref{e5}), while the long dashed line 
is obtained by exact numerical evaluation of 
Eq.~(\protect\ref{e2}) or Eq.~(\protect\ref{e3}). 
The short dashed line 
in (b) shows the $\ln(2A)$-contribution, while the dot-dashed 
line gives the $1/A^2$-contribution. 
The DM in 
(b) has been calculated on the particular representative node  
characterized by $\nu_{\rm x}=\nu_{\rm y}=\nu_{\rm z}=9$. } 
\label{f3} 
\end{figure}

Further analysis is based on the use of Stirling's approximation 
for $\nu_\alpha\gg 1$ which allows us to evaluate asymptotically 
both the internal finite sum $\Sigma_3$ in the $3D$ case and the 
external infinite sums over $k$ in all dimensions. For example, 
if all $\nu_\alpha \gg 1$, then keeping the leading term in the 
asymptotic expansions, we can write the following expression for 
the DM in all three dimensions: 
\begin{equation} 
\rho_{\nu_\alpha} \simeq 
(-1)^{\overline\nu}(2\pi)^{-D/2} 
\left(\prod_\alpha^D \nu_\alpha \right)^{-1/2} 
\exp\left[ -\sigma 
\lambda \right]~, 
\label{e4} 
\end{equation} 
where the inverse decay length, $\lambda(A,\nu_\alpha)$,  is 
given by 
\begin{equation} 
\lambda(A,\nu_\alpha) = \ln (2A) + \frac{\sigma\sigma_{-1}}{4A^2}+ 
\frac{\sigma_{\ln} }{\sigma}-\ln{\sigma}~, \label{e5} 
\end{equation} 
with $\sigma_{-1}\equiv \sum_\alpha^D \nu_\alpha^{-1}$ and 
$\sigma_{\ln} \equiv \sum_\alpha^D \nu_\alpha \ln{\nu_\alpha}$. 
As follows from Eqs.~(\ref{e4})-(\ref{e5}) and 
Figs.~\ref{f1}(a,c) and \ref{f2}(a,d),  the DM in this model of 
an insulator: (i) exponentially decays with the distance-related 
parameter $\sigma$; (ii) has a power-law prefactor which is 
proportional to $ \nu^{-D/2}$; (iii) is anisotropic;  and (iv) 
the inverse decay length of the DM depends on the energy 
parameters of the model. The behaviour with effective distance of 
$\ln|\rho_{\nu_\alpha}|$ and the effective inverse decay length, 
${\tilde\lambda} = -\ln|\rho_{\nu_\alpha}|/\sigma_1$, 
are 
shown in Figs.~\ref{f1}(a,c) for the $2D$ case and in 
Figs.~\ref{f2}(a,d) for the $3D$ case, respectively. The 
different points correspond to all the lattice nodes within a 
relatively wide solid (polar in $2D$) angle around the main 
diagonal. Examination of the results shows that the slowest 
decay of the DM occurs along the main diagonal, i.e. the 
$[111]$ direction in the $3D$ case. 
The analytical 
expression~(\ref{e5}) allows the different contributions to the 
effective decay length to be estimated. It is clearly seen in 
Figs.~\ref{f1}(c) and \ref{f2}(d) that the major $A$-dependent 
contribution comes from the $\ln(2A)$-term in Eq.~(\ref{e5}), 
while the role of the second term, $\sim 1/A^2$, in 
Eq.~(\ref{e5}) is not important at all, 
 especially in comparison with  the significant 
$A$-independent contribution from the last two terms in 
Eq.~(\ref{e5}) (see Fig.~\ref{f3}(b)). 
 
In the case when one of the indices, e.g. $\nu_{\rm x}\gg 
1$, but the other ones are small and finite ($D\ge 2$), 
\begin{equation} 
\rho_{\nu_\alpha} \simeq \frac{ (-1)^{\overline\nu}    } 
     { \sqrt{2\pi\nu_{\rm x}} } 
\exp \left[-\nu_{\rm x} \ln(2A) \right] \prod_{\alpha\ne {\rm x}} 
 J_{\nu_\alpha} \left( 
     \frac{\nu_{\rm x}}{A} 
\right)~, \label{e6} 
\end{equation} 
where $J$ stands for the Bessel function, the asymptotic form for 
which, with large argument \cite{Watson_66},    must be used in 
Eq.~(\ref{e6}). Bearing in mind that $J_{\nu_\alpha} (\nu_{\rm 
x}/A) \propto (\nu_{\rm x}/A)^{-1/2}$ for  $\nu_{\rm x}\gg 1$, it 
is easy to see that the power-law prefactor, $\propto \nu^{-D/2}$, 
is restored in Eq.~(\ref{e6}). The results for this particular 
case are presented in Figs.~\ref{f1}(b,d) and \ref{f2}(c,f). 
Again, the contribution of the term $\ln(2A)$ in the inverse 
decay length is dominant, and the inverse decay length  is now 
larger than for decay along the main diagonal (cf. e.g. 
Figs.~\ref{f2}(a) and \ref{f2}(c)). 
 
Finally, 
if  $\nu_{\rm x}, \nu_{\rm y} \gg 1$ and $\nu_{\rm z}$ is finite (in $3D$),  
%
%
\begin{eqnarray} 
\rho_{\nu_\alpha} 
&\simeq& 
(-1)^{\overline\nu} \sqrt{ 
      \frac 
           { \nu_+   } 
           {2\pi \nu_{\rm x}\nu_{\rm y}  } 
      } 
\exp\left[ 
     -\nu_+ 
      \left(1+ \frac{ \nu_- } 
               { 2\nu_+ } 
                 \ln(\nu_{\rm x}/\nu_{\rm y}) 
      \right) 
    \right] 
\nonumber \\ 
& & 
J_{\nu_{\rm z} } \left[ 
      \frac{\nu_+ }{A} 
\right] J_{\nu_+ } \left[ 
      \frac{ \nu_+^2 } 
           { \sqrt{ \nu_{\rm x} \nu_{\rm y} } A } 
\right]~, 
\label{e7} 
\end{eqnarray} 
%
%
where $ \nu_{\pm} = \nu_{\rm x}\pm \nu_{\rm y}$ and, for both 
Bessel functions, known asymptotic expressions (for the argument 
in $ J_{\nu_{\rm z} }$ and for the argument and index in $ J_{ 
\nu_+  }$) must be used \cite{Watson_66}. These asymptotics for 
both Bessel functions restore the $\nu^{-3/2}$ prefactor for the 
DM in the same manner as for Eq.~(\ref{e6}). The energy 
($A$)-dependent contribution to $\lambda$ (see 
Fig.~\ref{f2}(e)), in this case, comes  mainly from the 
asymptotic expression for $ J_{ \nu_+  }$ which is of exponential 
type if $\nu_+/ (A\sqrt{\nu_+}) <1$ (and of $\cos$-type 
otherwise) \cite{Watson_66}. 
 
Another important point to discuss is the dependence of the 
typical inverse decay length, $\lambda$,  on the energy parameters 
of the model. 
The only energy-dependent parameter in this  model 
is the parameter $A$ 
which is simply related to the spectral characteristics of the model, 
$A=\Delta\varepsilon/4t_{12} = D\Delta E_{\rm opt}/(\beta\Delta E_{\rm b})$, where 
$ \Delta E_{\rm b}=4Dt$ is the band width, 
$ \Delta E_{\rm opt} =\Delta\varepsilon$ is 
the optical (direct) band gap 
and the ratio of hopping integrals, $\beta = t_{12}/t$, 
which can be connected to the thermal 
gap, 
$\Delta E_{\rm th}= [\Delta E_{\rm opt}^2+ 
(\beta \Delta E_{\rm b})^2]^{1/2}-\Delta 
E_{\rm b}$.  
The spectral parameters 
($\Delta E_{\rm b}$, $ \Delta E_{\rm opt}$ and 
$ \Delta E_{\rm th} $) are not independent values 
because the hopping integrals $t_{\mu\mu'}$  
 in the Hamiltonian~(\ref{e1}) can depend 
on the orbital energies $\varepsilon_\mu$  
and obtaining such a dependence, 
e.g. $t_{12}(\Delta\varepsilon)$, requires more accurate analysis 
which is not necessary for our approach 
(that is why we plot  $\lambda$ vs $A$ 
but not  the direct gap width $\Delta\varepsilon$ 
in Fig.~\ref{f3}). 
However, the limiting case,  $\Delta\varepsilon \to 0$,  
can be investigated, at least numerically (analytical 
results are valid only for $A>D$) under the reasonable assumption that $t_{12}\ne 0$ 
as $\Delta\varepsilon \to 0$, and the results are presented in 
Fig.~\ref{f3}(a), confirming the linear scaling of the inverse decay 
length, $ \lambda \propto A \propto \Delta\varepsilon$ 
for $A\to 0$ \cite{IsmailBeigi_99}. 

This model can be analysed analytically 
 only for a s.c. lattice and 
thus, strictly speaking, we stick to  
the atomic orbital representation. 
However, numerics show that the results 
(exponential decay of the density matrix and variation of the 
decay length with parameters of the Hamiltonian) 
are qualitatively the same for different underlying 
lattices. 
This  is not surprising because the asymptotic behaviour  
at large distances hardly depends on the local structural details 
(the local features of the lattice can influence the value of $\lambda$ 
by a factor of $\sim 2$; see Figs.~\ref{f1}-\ref{f2}). 
Thus, we believe that our results, at least qualitatively, are general. 
Indeed, the inverse decay length decreases with decreasing $A$ 
(see Fig.~\ref{f3}), thus reflecting 
the known delocalization tendency of the DM with increasing 
metallicity (in the bonding/antibonding representation 
\cite{Harrison_80} $A \simeq V_2/V_1$,  being the ratio of the 
bonding-antibonding splitting energy and the metallic bandwidth energy). 
Moreover, the analytical expressions for the decay 
length give correct order-of-magnitude estimates 
of $\lambda$ for Si and C  
\cite{dadnote}. 
 
To conclude, we have derived analytic asymptotic expressions  for 
the spatial decay of the density matrix decay in insulators. The 
results have been obtained for a tight-binding Hamiltonian 
defined on the s.c. lattice having two orthogonal bare 
states on each node. 
Two additional assumptions, namely 
symmetric bands (equal in-band transfer integrals) and relatively 
weak intraband hybridization, allow us to derive exact asymptotic 
results for the density matrix in all three dimensions. The main 
features of the analytic solution are: exponential spatial decay 
of the density matrix, anisotropy of the decay length, the 
dependence of the decay length on the energy parameters of the 
model, and the existence of a power-law prefactor ($\propto 
r^{-D/2}$). All the analytic asymptotic results have been 
supported by precise numerical solutions of the problem. 
 
We are grateful to Peter Paule for giving us access to the finite 
hypergeometric summation codes. We thank L. He and D. Vanderbilt 
for helpful comments on the manuscript. DAD acknowledges support 
from the NSF under grants DMR 0081006 and DMR 0074624.


\begin{thebibliography}{10}
\bibitem{dadnote} 
Using the relation  $A = V_2/V_1$ 
(relevant  for the 
bonding/antibonding representation)
 and the values of parameters 
given on p. 90 of Ref.~\protect\cite{Harrison_80},  
we can estimate $\lambda 
\approx ln(2A) = 1.22$ ($1.87$) for diamond Si (C), which are 
rather close to the values   $\lambda \simeq 1.15$  ($2.06$) 
obtain by realistic calculations in Ref.~\protect\cite{Zhang_01}. 
These estimates are coarse since the diamond lattice does 
not have bond orbitals on a s.c. lattice and we use only the 
simplest approximation for $\lambda$.
\end{thebibliography}

\end{document}